\documentclass[12pt,reqno]{amsart}
\usepackage{latexsym}
\usepackage{amsmath}
\usepackage{amscd}
\usepackage{amssymb}
\usepackage{mathrsfs}

\usepackage{enumerate}

\usepackage{graphicx}
\usepackage{subfigure}
\usepackage{url}

\begin{document}

\title{The stochastic integrable AKNS hierarchy}

\author{Alexis Arnaudon}

\address{Department of Mathematics, Imperial College, London SW7 2AZ, UK}

\email{alexis.arnaudon@imperial.ac.uk}
\maketitle

\begin{abstract}
	We derive a stochastic AKNS hierarchy using geometrical methods. The integrability is shown via a stochastic zero curvature relation associated with a stochastic isospectral problem.   
	We expose some of the stochastic integrable partial differential equations which extend the stochastic KdV equation discovered by M. Wadati in 1983 for all the AKNS flows. 
	We also show how to find stochastic solitons from the stochastic evolution of the scattering data of the stochastic IST. 
	We finally expose some properties of these equations and also briefly study a stochastic Camassa-Holm equation which reduces to a stochastic Hamiltonian system of peakons. 
\end{abstract}


\section{Introduction}

Completely integrable partial differential equations play a major role in mathematics as well as in physics. 
The structure of these equations and their solutions were and will always be interesting mathematical problems. 
On top of this a vast majority of these equations naturally appear in a wide range of physical problems, from shallow water waves to optical fibres.
One has to be careful with such an affirmation because complete integrability only appears in physics after very specific assumptions and simplifications of the physical model under consideration.
Any realistic modelling cannot be completely integrable in the strict mathematical sense. 
Nevertheless the characteristic solutions of integrable systems that are solitons may be observed in nature, with of course only a finite life time. 
During this time it is common to approximate the physical model by a completely integrable system in order to access the whole range of powerful mathematical tools developed for these equations. 
If one still wants to study a more realistic equation, perturbations of integrable systems can be considered. 
This has been a vast area of study and we only refer to \cite{bass1988dynamics,kivshar1989dynamics,abdullaev1989dynamical,konotop1994nonlinear} for reviews of this subject. 
The perturbations are usually given by an external potential or forcing, high order nonlinearities, diffusions, high order dispersions or random fluctuations. 
The complete integrability is lost in almost every cases of above mentioned perturbations and it is a difficult problem to find perturbations which retain integrability. 
This paper is devoted to a very specific random perturbations of integrable systems that preserves their complete integrability.

As already discussed above the link with physical models is of course not straightforward but, upon relaxing assumptions which would break the complete integrability, such connection can be made. 
Our main assumption for this work is that the Brownian motion, or Wiener process, will be strictly temporal. 
Recently some works have been devoted to the study of stochastic NLS or KdV equations with a space-time dependent noise. 
We refer to \cite{debouard1998stochastic,debouard1999stochastic} or the book \cite{da2014stochastic} for theoretical studies and to \cite{debussche1999numerical,debussche2002numerical} for numerical experiments. 
Note that there also exists a noisy second order dispersion for the NLS equation with a temporal noise, studied for example in \cite{debouard2010nonlinear}.  

Our interest is in the derivation of stochastic integrable equations, an idea which began with Wadati in \cite{wadati1983stochastic,wadati1984stochastic} (see also the book \cite{konotop1994nonlinear}).  
Note that a few years later \cite{blaszak1986kdv,blaszak1986mkdv} extended the stochastic KdV to the entire KdV and mKdV hierarchies. 
We also refer to \cite{herman1990stochastic} for more works on the stochastic KdV equation and we note that the diffusion of solitons predicted in \cite{wadati1983stochastic} has later been numerically tested for example by \cite{scalerandi1998korteweg}. 
These works only considered the simple case of an additive temporal noise that  we will recover in our more general framework as one of the possible types of stochastic deformations. 

Adding noise in a systematic way to integrable systems while preserving their integrability must be done at an abstract level in the theory of integrable systems in such a way that we reduce the number of possibilities for adding noise. 
The theory that we will use here is based on the original construction of the AKNS hierarchy of \cite{ablowitz1974inverse}, which can be understood as a common framework for the KdV, mKdV or NLS equations. 
We will use the geometrical framework of this hierarchy developed in \cite{arnaudon2015lagrangian} and in particular the concept of a Lagrangian reduction for deriving the AKNS hierarchy.
This method already led to another type of deformations of integrable systems via the Sobolev $H^1$ norm in \cite{arnaudon2015lagrangian}. 
We will then need a geometrical method for adding noise such that this geometrical framework is preserved and thus the integrability of the corresponding PDEs. 

Over the years several authors proposed different methods for adding noise to a mechanical system with the aim of being compatible with the geometrical properties of the deterministic system.  
The first theory was developed by Bismut in \cite{bismut1982mecanique} and inspired many other later works. 
We refer to the quantum mechanical approach of \cite{zambrini2012research,zambrini2013stochastic}, the reduction by symmetry of \cite{arnaudon2014stochastic,chen2015constrained} or a theory of stochastic Hamiltonian systems of \cite{lazaro2008stochastic}. 
Another approach was recently proposed in \cite{holm2015variational} in the context of fluid dynamics and then applied to finite dimensional mechanical systems in \cite{arnaudon2015fokker}. 
We will follow this approach here as it perfectly fits our Lagrangian reduction framework of integrable systems and produces stochastic PDEs. 
The other approach compatible with the Lagrangian reduction by symmetry is \cite{arnaudon2014stochastic} but would instead produce dissipative terms in the PDE rather than stochastic processes. 
We refer for example to the recent derivation of the dissipative MHD equation in \cite{chen2015constrained} or the viscous Camassa-Holm equation using this theory in \cite{cruzeiro2015stochastic}.

We want to mention that the framework of \cite{holm2015variational} applied to finite dimensional mechanical systems in \cite{arnaudon2015fokker} also led to the discovery of a few integrable stochastic mechanical systems such as the Lagrange top or the free rigid body.
Another application of \cite{holm2015variational} for finite dimensional systems was done for the study of a stochastic dynamics of peakons in \cite{holm2015soliton}. 
Unfortunately the reduction from a stochastic CH equation is not possible for obvious reasons, but we will show such reduction from the integrable stochastic CH equation to a different stochastic peakon system.  

The main result of this paper consists of the combination of the Lagrangian reduction and the stochastic deformation of this system. 
This combination led to the classification of stochastic integrable systems and the development of the stochastic inverse scattering transform. 
The structure of the stochastic equation turn out to be rather simple as the noise term is always proportional to a flow of the same integrable hierarchy. 
As our method is rather general we expect that the stochastic equations derived in this work are the only instances of integrable stochastic NLS, KdV and mKdV equations. 
Recently \cite{kedziora2015integrable} obtain similar results only for the NLS reduction of the AKNS hierarchy and a noise replaced by a deterministic time dependent function. 
Our results can be recovered here if one replaces the Weiner process $W_t$ by a deterministic function $f(t)$. 

\subsection*{Plan of the paper} 
We first review the theoretical framework of the Lagrangian reduction for the AKNS hierarchy in Sec.~\ref{lagrangian}, then we will describe the inclusion of noise into this type of Lagrangian system in Sec.~\ref{clebsch}. 
The stochastic AKNS hierarchy is then derived in Sec.~\ref{SAKNS} and summarized in table \ref{tab:classification}. 
The stochastic inverse scattering transform is then presented in Sec.~\ref{SIST} and ends the theoretical section.
The next Sec.~\ref{SPDE} is devoted to the study of some of the stochastic integrable equations, with a short paragraph on the stochastic CH equation.

\section{Theoretical framework}
We first expose the main tools needed for a Lagrangian reduction of integrable systems and the injection of noise via a constraint variational principle. 
The main reference for the Lagrangian reduction in this context is \cite{arnaudon2015lagrangian} and for the stochastic constraint variational principle \cite{holm2015variational}. 

\subsection{Lagrangian reduction}\label{lagrangian}

\subsubsection{Loop groups and algebras}
Hierarchy of integrable systems can be understood as a reduction by symmetry where the group of symmetry, which is also the configuration space, is an infinite dimensional Lie group, the so-called loop groups. 
This group consists of maps from the circle $S^1$ into a finite dimensional Lie group $G$. 
We will always denote by $\lambda$ the coordinate on $S^1$, which will later be understood as the eigenvalue of an isospectral problem.
Although this type of group is very general we will focus here on the simplest case, namely when the group is $G= SL(2)$.
As we will see, the structure of this group will naturally produce the AKNS hierarchy of \cite{ablowitz1974inverse}. 
In fact, we only require the group to be semi-simple, in order to identify its Lie algebra and its dual via a non-degenerate pairing given by the Killing form. 
This group has a natural Lie algebra, given by maps from $S^1$ to $\mathfrak g$, the Lie algebra of $G$.
We will denote the loop algebra by $L\mathfrak g$ and the loop group by $LG$. 
The loop algebra also carries a natural pairing, via integration along $S^1$ and the trace pairing of the Lie algebra. 

\subsubsection{Central extension and cocycle}

In order to derive 1+1 PDEs, we must introduce a space variable, denoted by $x\in \mathbb R$. 
We will also make the assumption of vanishing boundary conditions at $\pm \infty$, such that we can freely integrate by parts. 
The loop group or loop algebra elements we will consider will thus depend on the time $t\in \mathbb R^+$, the space $x$ and the loop parameter $\lambda$. 
We will shortly show how to recast the integrable hierarchies into the standard theory of reduction  in classical mechanics but before we must explain how to include the $x$ dependence of the dynamical equations in this framework. 
This is done with the help of a central extension of the loop group, characterized by a cocycle. 
The central extension consists of appending a vector space to the group with the property that only the action of the group on itself will produce an element in this vector space, called cocycle.  
In our case the vector space will just be $V= \mathbb R$, the cocycle is $C(g,h)= \int g\partial_xh dx$ for $g,h\in LG$. 
We will denote the loop group with central extension by $\widetilde {LG}$ and its Lie algebra $\widetilde{L\mathfrak{g}}$. 
The group action is thus $(g,a)\cdot (h ,b)= (g\cdot h, C(g,h))$, for $(g,a),(h,b)\in LG\times V$ and $g\cdot h $ is the group action of $G$, point wise in $\lambda$. 
In order for this action to be valid, the cocycle $C$ has to satisfy a cocycle identity, see \cite{marsden2006book} for more details. 
In our case of a derivative cocycle this identity holds and the cocycle also directly reduces to the Lie algebra to read $c(\xi,\eta) = \int \xi\partial_x \eta dx$ for $\xi,\eta\in L\mathfrak g$.
In this case, one can check that the adjoint and coadjoint actions are given for $(\xi,a),(\eta,b)\in L\mathfrak g$ and $(\mu,m)\in L\mathfrak g^*$, by
\begin{align}
	\mathrm{ad}_{(\xi,a)}(\eta,b)&=  ([\xi,\eta],c(\xi,\eta))\label{adj}\quad \mathrm{and}\\
	\mathrm{ad}^*_{(\xi,a)}(\mu,m)&= ([\xi,\mu] + mc(\xi,\cdot),0)\label{coadj},
\end{align}
where the adjoint and coadjoint coincide owing to the non-degenerate pairing on the Lie algebra. 

\subsubsection{Structure of the hierarchy}
In order to understand the entire integrable hierarchies such as the AKNS hierarchy, the loop group is a fundamental ingredient but another concept is needed. 
The hierarchies come as an infinite number of PDEs, all sharing the same conserved quantities and which can all be recovered from just the recursion operator. 
We will not show here the construction of this operator, which is based on the existence of a bi-Hamiltonian structure of the integrable PDE, but rather focus on the zero curvature relation (ZCR). 
This relation is fundamental in the theory of the inverse scattering transform (IST) which is used to find explicit solutions of integrable PDEs. 
The ZCR for the PDE of a same integrable hierarchies are linked between each other and perfectly fit the framework of loop group and reduction by symmetry with central extension. 
Indeed, the spectral parameter of the IST is $\lambda$, the loop parameter and the Euler-Poincar\'e, or Lie-Poisson equations coming from reduction by symmetry will be corresponds to the ZCR, as we will shortly demonstrate.  
We first have to construct the hierarchy, form the loop algebra point of view. 
For this, we define a generic loop algebra element  $M^{(\infty)}= \sum_{i=0}^\infty M_i\lambda^{-i}\in L\mathfrak g$ where $M_i\in \mathfrak g\, \forall i$. 
Then, out of this $M^{(\infty)}$ we construct other loop algebra elements as $M^{(i)} =P_+(\lambda^iM^{(\infty)})\, \forall i$, where $P_+$ is the projection onto the positive powers of $\lambda$.
This projection is linked to the decomposition of the loop algebra into subalgebras with positive and negative powers of $\lambda$, see \cite{flaschka1983kac,newell1985solitons}.
Each of the $M^{(i)}$ loop algebra elements will be associated to a given space or time direction, in the following sense. 
We extend the two dimensional space-time of the 1+1 PDE to an infinite dimensional space time with coordinates $a= (a_1,a_2,\ldots)\in \mathbb R^\infty$. 
The 1+1 spacetime is thus a slice of this infinite dimensional space, denoted $N_{ij} \subset \mathbb R^\infty$ with coordinates $(a_i,a_j)$. 
We then construct a one form on this space with values in the loop algebra from the $M^{(i)}$ as $\mathbf M = \sum_i M^{(i)}d a_i$. 
This one form is in fact a connection on the bundle structure $L\mathfrak g\times \mathbb R^\infty \to \mathbb R^\infty$, and we require its curvature to vanish, written $d\mathbf M + [\mathbf M,\mathbf M] = 0 $. 
In components, the relation decomposes into an infinite number of equations, indexed by two integers $i,j$, written 
\begin{align}
	\partial_{a_i} M^{(j)} -\partial_{a_j} M^{(i)}+ [ M^{(j)},M^{(i)}] = 0,\quad \forall i,j.
	\label{ZCR}
\end{align}
As we said this relation holds on every slice $N_{ij}$ and in particular for the slice $N_{i\infty}$ and simplifies to  
\begin{align}
	\partial_{a_i} M^{(\infty)} + [ M^{(\infty)},M^{(i)}] = 0. 
	\label{ZCR-leg}
\end{align}
We now have a relation between the $i$ fields $M_j$ with $j\leq i$ contained in $M^{(i)}$ with all the other fields contained in $M^{(\infty)}$. 
This relation is fundamental in our construction as it allows us to express the infinite number of  Lie algebra elements $M_i$ in term of only a finite number of them. 

\subsubsection{Euler-Poincar\'e and Lie-Poisson equations}

We now briefly show how to interpret the ZCR on a slice $N_{ij}$ as either a Hamiltonian system or a Lagrangian system. 
The subtlety come from the interpretation of the coordinates in $N_{ij}$ with respect to the relation \eqref{ZCR-leg}.
Indeed, if we choose $i,j$ such that $i>j$, set $t:= a_i,\, x:=a_j$ and use the coordinate $a_i$ to compute all the $M^{(j)}$ from \eqref{ZCR-leg} as a function of $M:= M^{(i)}$, and in particular $L:= M^{(j)} (M)$, we can make sense of a Lagrangian $l(M)= \int M\cdot L(M) dx$. 
Then, the Euler-Poincar\'e equations associated to this Lagrangian, with the co-adjoint action defined in \eqref{coadj}, reads
\begin{align}
        \partial_t \frac{\delta l(M)}{\delta M} - \partial_x M + \left [M,\frac{\delta l(M)}{\delta M} \right] = 0,	
	\label{EP}
\end{align}
where the fact that $\frac{\delta l(M)}{\delta M} = L$ makes the connection with \eqref{ZCR} clear. 
Note that because we chose $i> j$ the order of the polynomial for $M$ is bigger than for $L$, thus the Euler-Poincar\'e equation \eqref{EP} is in fact a system of equation for the fields in $M$ (which are also in $L$ from the relation \eqref{ZCR-leg}). 
This equation is a direct consequence of the reduction by symmetry from the loop group to the loop algebra, see \cite{arnaudon2015lagrangian} and references therein for more details. 

If we now follow the same reasoning but with $x:=a_j$ for the calculation of the $M^{(i)}$ with \eqref{ZCR-leg} we will arrive at $L:= M^{(j)}$ being the fundamental field, and thus $M(L):= M^{(i)}$. 
We can then associate a Hamiltonian $h(L)= \int L\cdot M(L)dx$, where the corresponding Lie-Poisson equation with cocycle will have the same form as \eqref{EP}, but written with these different variables. 
We do not write the Lie-Poisson equation here but it can be found in \cite{arnaudon2015lagrangian}.
Note that the number of equations form the Lie-Poisson equation is smaller than for the Euler-Poincar\'e equation, as $L$ contains less independent fields than $M$ and is the fundamental variable.  
The interpretation of \eqref{ZCR-leg} is now clear: it corresponds in the classical mechanical setting to an equation for finding the moment of inertia.

The last step is to find the relation between these two formalisms, or the Legendre transformation between the Lagrangian and the Hamiltonian equations. 
This can easily be seen after noticing that the extra fields in $M$ compared to $L$ can be expressed in terms of the fields in $L$ with the help of some of the equations in the Euler-Poincar\'e equation \eqref{EP}. 
We will only treat here the simple case of $\mathfrak g= \mathfrak {sl}(2)$.
This case is the standard AKNS hierarchy, that we will stochastically deformed here.
We briefly recall standard facts about $\mathfrak {sl}(2)$, which consists of traceless two by two matrices, with the following standard basis
\begin{align}
	\mathbf h= 
	\begin{bmatrix}
		1 & 0\\
		0 & -1
	\end{bmatrix},\quad 
	\mathbf e= 
	\begin{bmatrix}
		0 & 1\\
		0 & 0 
	\end{bmatrix}\quad \mathrm{and} \quad 
	\mathbf f= 
	\begin{bmatrix}
		0 & 0 \\
		1 & 0 
	\end{bmatrix},
\end{align}
where $\mathbf h$ spans the one dimensional Cartan subalgebra. 
We will use the notation, for a generic $U\in \mathfrak{sl}(2)$,  $U^\perp:= U_\mathbf{e} + U_\mathbf{f}$ and $U^\|= M_\mathbf{h}$. 

Returning to our derivation of the Legendre transform, we have that for example if $i=2,j=1$ we would start with $M= \lambda^2A + \lambda U + V$ and find $L(M) = \lambda A + U^\perp$ from the relation \eqref{ZCR-leg}, where $A=ih$ is a constant in the Cartan subalgebra of $\mathfrak g$ and $\perp$ projects away the Cartan subalgebra part of $U$.  
Then, the Euler-Poincar\'e equations \eqref{EP} contain the extra equation $\partial_s U = [V,U^\perp]$ which can be inverted to give $V(U)$ and thus express everything in term of $U$ only, or $L$ only. 
Having done this, the equation becomes the Lie-Poisson equation for $L$, which could have been directly found in the Hamiltonian setting.  

From there, one can derive the AKNS hierarchy, which contains the well-known NLS, KdV, mKdV equations, be specifying a particular form for $U$, see \cite{arnaudon2015lagrangian} or below. 
This Lagrangian framework was already used for developing a deformation of this integrable hierarchy by using a Sobolev $H^1$ norm in the Lagrangian, instead of the $L^2$ norm. 
This led to the CH and mCH equations as well as a deformation of the NLS equation, studied in \cite{arnaudon2015deformation}. 
We will not follow this track here, but use this Lagrangian interpretation to add a particular type of noise while preserving the geometric structure of the equations. 

\subsection{Stochastic constraint variational principle}\label{clebsch}

We are now ready to introduce noise into the Lagrangian framework of integrable hierarchies, following the method developed by \cite{holm2015variational}.
The main idea for introducing noise into a mechanical system while preserving as much geometry as possible is based on the remarkable fact that the Euler-Poincar\'e equations can be, independently from the reduction method, recovered from a constraint variational principle, see \cite{gay2011clebsch} and references therein. 
This correspondence was used to inject noise from the associated constraint variational principle into the equation of motion. 
The resulting stochastic process is called the stochastic Euler-Poincar\'e equation.  
Note that this noise can also be understood as coming from a noisy version of the reconstruction relation (see for example \cite{arnaudon2015fokker}), but because the reconstruction of the solution on the group is not of interest here we will stick to the stochastic constraint approach. 

In this work we restrict ourselves to the simplest type of noise, that is a temporal Brownian motion denoted by a Wiener process $W_t$.  
Throughout this paper we will mainly use Stratonovich integrals, denoted by $\circ dW_t$ such that we can use standard rules of calculus. 
The It\'o formulation will be simply denoted by $dW_t$.  
We refer for example to the first section of \cite{chen2015constrained} for the definitions and properties of this noise but we will not use the manifold machinery of stochastic processes here as we will always work on the Lie algebra, or vector space level. 

The injection of noise into the integrable hierarchies goes as follow. 
We first start with the constraint variational principle, written for a time $t_i$ and $M:= M^{(i)}$ as 
\begin{equation}
	\begin{split}
		S= \int \left ( l(M)+ \left \langle (p,\mu),\partial_{t_i}\left (q,a\right ) +\mathrm{ad}_{(M,\xi)}(q,a)\right \rangle \right)dt_i
	+ \int \Phi(q,p)\circ dW_{t_i}.
	\end{split}
	\label{clebsch-sto}
\end{equation}
where $(p,\mu)\in T^*\widetilde{LG}$ is the so-called Clebsch variables, or Lagrange multipliers used to enforce the constraint that $(q,a)\in T\widetilde {LG}$ evolves advected by $(M,\xi)\in \widetilde{L\mathfrak g }$.
Recall that $\widetilde{ LG}$ is the loop group with central extension, therefore the element $a$ is a polynomial in $\lambda$ with the same order as $q$ but with values in $\mathbb R$.  
The function $\Phi$ is the stochastic potential which introduces a noise into this variational principle. 
The exact form of $\Phi$ will be determined later such that the stochastic Euler-Poincar\'e equation will  correspond to a stochastic ZCR but we can already assume that $\Phi$ does not depend on the centre of the loop algebra.   
We first compute the variations of \eqref{clebsch-sto} to find, using \eqref{adj} and \eqref{coadj}, the set of equations 
\begin{equation}
	\begin{split}
\frac{\delta l}{\delta M} &=\mathrm{ad}^*_qp+\mu c(q,\cdot)=: L\\
(d_{t_i} p,d_{t_i} \mu) &= \mathrm{ad}^*_{(M,\xi)}(p,\mu)dt+\frac{\delta \Phi}{\delta q}\circ dW_{t_i}\\
(d_{t_i} q,d_{t_i} a) &= - \mathrm{ad}_{(M,\xi)}(q,a)dt -\frac{\delta \Phi}{\delta p}\circ dW_{t_i}.
	\end{split}
	\label{variations}
\end{equation}
Note that the equation do not produce any dynamics in the centre, thus $\mu$ and $a$ are arbitrary constants. 
In order to go on with the computation and get rid of the extra $q$ and $p$ variables we must make an extra assumption on $\Phi$. 
The most natural thing we can do is to let $\Phi= \Phi(L)$ as $L$ contains the only field which will interest us for a forthcoming stochastic integrable PDE. 
Then, after having defined the Stratonovich stochastic process
\begin{equation}
	dX_i := Mdt_i+\frac{\delta \Phi}{\delta L}\circ dW_{t_i},
	\label{dX}
\end{equation}
the system \eqref{variations} can be rewritten by computing $dL$ and using the Jacobi identity to give the stochastic Euler-Poincar\'e equation 
\begin{equation}
	d \frac{\delta l(M)}{\delta M}= \mathrm{ad}^*_{dX_i}\frac{\delta l(M)}{\delta M}+c(dX_i,\cdot).
	\label{EP-sto}
\end{equation}
Equivalently, using the derivative cocycle and the pairing of the semi-simple Lie algebra $\mathfrak g$ we arrive at
\begin{equation}
	d L-\partial_x dX_i+ [L,dX_i]= 0.
	\label{ZCR-sto}
\end{equation}
Note that the spatial derivative is well define as the noise $dW_{t_i}$ does only depend on the time and will not modify the regularity in space of the solutions. 
This particular type of noise thus affects the velocity $M$ of the system with a stochastic drift as seen from Eq.~\eqref{dX} and keep the geometric structure of the stochastic equations as seen from Eq.~\eqref{ZCR-sto}. 
A standard additive or multiplicative noise at this level of this theory would surely not preserve any geometry nor any integrability.  

\subsection{The stochastic AKNS hierarchy}\label{SAKNS}

From the stochastic EP equation \eqref{ZCR-sto}, we can now calculate the Legendre transform in order to obtain the corresponding stochastic PDEs.  
Note that, as compared to the deterministic case, the stochastic potential $\Phi(L)$ has to be determined. 
In fact we just have to find $\Xi:= \frac{\delta \Phi(L)}{\delta L}$, thus the corresponding stochastic velocity drift.
By rewriting  the stochastic Euler-Poincar\'e equation \eqref{ZCR-sto} explicitly in term of $\Xi$, we obtain 
\begin{equation}
	d L = (\partial_xM^{(i)}- [L,M^{(i)}])dt_i + (\partial_x\Xi- [L,\Xi])\circ dW_{t_i}.
\end{equation}
It is then clear that $\Xi$ will be find via the same Legendre transformation procedure that is used to express $M^{(i)}$ in term of $L$. 
We have here another degree of freedom, as we can choose the order of the polynomial for the loop algebra element $\Xi$. 
We will denote it with the index $k$, and one can easily be convinced that $\Xi(L)= M^{(k)}(L)$, where $M^{(k)}$ is computed via the Legendre transform in the deterministic case. 
Hence, we will call $k$ the stochastic time. 
The stochastic deformation of the hierarchy will thus have three choices of space time indices: $j$ for the space, $i$ for the time and $k$ for the stochastic time. 
The flow thus lives on the slice $N_{ij}$ but is stochastically deformed in the direction $N_{ik}$. 
We will now illustrate this stochastic deformation with the most important integrable hierarchy which is the AKNS hierarchy with $j=1$. 

Note first that the solution of \eqref{ZCR-leg} is simple and always give $L= \lambda A + M_1^\perp$ where $A= i\mathbf h$ because we stick to our choice of $x:= t_1$. 
The choice of complex $A$ is necessary to obtain the NLS equation. 
From here, we have to compute the Legendre transform, which consist in finding $M_i, \forall i$ as a function of $U:= M_1$ by using the Euler-Poincar\'e equation \eqref{ZCR-sto}.
We therefore only need to compute the $M^{(i)}$ that we are interested in and combine them to obtain the corresponding stochastic integrable PDE. 
This calculation can be found in \cite{arnaudon2015lagrangian} and gives the usual AKNS hierarchy flows, provided that $U$ has a specific form. 
For $U^\perp = q\mathbf e + r \mathbf f$, the reduction $q=u, r= \overline u$ gives the NLS flow for $t_2$, $q= r= u$ the mKdV equation for $t_3$ and for the same time $q=1, r=u$ gives the KdV equation. 
Now, associated with these standard flow we can choose another time for the stochastic deformation, but then, depending on this choice, not every reductions will still gives integrable equations.  
This comes from the fact that it is not possible to mix different flows with different reductions, for example NLS and KdV.  
We give in table \ref{tab:classification} the first possible reductions associated with the flows which gives integrable stochastic equations.

\begingroup
\renewcommand{\arraystretch}{1.2}
\begin{table*}
	\centering
	\caption{This table presents a classification of the stochastic deformation of the first integrable equations of the AKNS hierarchy, for the time flow $t_i$ and the stochastic time $t_k$.  The second column displays the reductions of the matrix $U$ in order to arrive at the equation of the third column. }
	\begin{tabular}{|c|c|c|}
		\hline
		$(i,k)$, Sec. & Reduction & Stochastic PDE \\ \hline \hline
		(2,1), \ref{SNLSI-section}& $q= u,r= \overline u 	$ & $du =  i\left (u_{xx} +2u|u|^2\right )dt +  u_x\circ dW_t $\\ \hline
		(2,2), \ref{SNLSII-section}& $q= u,r= \overline u 	$ & $du =  i\left (u_{xx} +2u|u|^2\right )\left (dt + \circ dW_t\right ) $\\ \hline
		(2,3), \ref{SNLSIII-section} & $q= u,r= \overline u 	$ & $du =  i\left (u_{xx} +2u|u|^2\right )dt + \left (u_{xxx} +6|u|^2u_x\right )\circ dW_t $\\ \hline 
		(3,1), \ref{SNLSI-section} & $q= u,r=  u^n$, $n=0,1$  & $du =  \left (u_{xxx} +6u^{n+1}u_x\right )dt +  u_x\circ dW_t $\\ \hline
		(3,2), \ref{SKdVII-section} & $q= u,r=  \overline u$ & $du =  \left (u_{xxx} +6|u|^2u_x\right )dt + i\left (u_{xx} +2u|u|^2\right )\circ dW_t $\\ \hline
		(3,3), \ref{SNLSII-section} & $q= u,r=  u^n$, $n=0,1$  & $du =  \left (u_{xxx} +6u^{n+1}u_x\right )(dt + \circ dW_t) $ \\ \hline
	\end{tabular}
	\label{tab:classification}
\end{table*}
\endgroup

\subsubsection{Stochastic bi-Hamiltonian structures}
Before entering into the details of the stochastic inverse scattering transform associated with the stochastic ZCRs derived before we want to understand the stochastic bi-Hamiltonian structures of these equations. 
As we already saw earlier, the present method for adding noise preserves the geometric structure of the hierarchy.  
In particular it preserves the Hamiltonian structures and the recursion operators of the hierarchies.
The only modification is in the definition of the Hamiltonian, the stochastic part of the equation has his own Hamiltonian, called stochastic potential. 
The bi-Hamiltonian system is then for the form
\begin{align}
	d u& = J\frac{\delta h_i}{\delta u} dt_i + J\frac{\delta h_k}{\delta u} \circ dW_{t_i}  = K\frac{\delta h_{i+1}}{\delta u} dt_{i+1} + K\frac{\delta h_{k+1}}{\delta u} \circ dW_{t_{i+1}},
\end{align}
for the two Hamiltonian structures $J$ and $K$, Hamiltonian $h_i$ and stochastic potential $h_k$. 
Note that the stochastic potential is another conserved quantity of the hierarchy with index $k$ independent of $i$. 
The notation for $u$ should be adapted to the equation in consideration, for example the stochastic NLS equation will have $u\to  (u, \overline u)$ and $J$ and $K$ would be the usual NLS Hamiltonian structures. 
It is clear from here that $h_i$ are all quantities exactly conserved by the corresponding stochastic integrable equations.  

\subsubsection{Stochastic inverse scattering transform}\label{SIST}

In \cite{arnaudon2015lagrangian} the Euler-Poincar\'e equation was shown to correspond to the ZCR used in the IST. 
The same is true here with the only difference that the Euler-Poincar\'e equation and the ZCR are stochastic.
This ZCR corresponds to a stochastic isospectral problem that will be studied in this section.
Note that this is a similar method used by \cite{wadati1983stochastic} to find the solution of his stochastic KdV equation.

The stochastic ZCR \eqref{ZCR-sto} is equivalent to the compatibility condition $\partial_x d\Psi = d\partial_x \Psi$ for a wavefunction $\Psi(x,t,\lambda)$ evolving in time and space as 
\begin{align}
	d\Psi &= M\Psi dt  + \Xi\Psi \circ dW\label{scatt-t}\\
	\partial_x \Psi &=L\Psi.\label{scatt-x}
\end{align}
If $\Xi=0$ the standard isospectral problem is recovered and in our case the time evolution of wavefunction $\Psi$ is a stochastic process. 
We will restrict ourselves to the IST with the NLS reduction of $U$, explain for example in \cite{ablowitz1981solitons,ablowitz2004discrete}. 
Although the scattering data and the construction of the solution will be the same as in the standard theory, the time evolution of the scattering data is modified and becomes stochastic. 

We first briefly recall the definition of the scattering data associated to the scattering problem \eqref{scatt-x}. 
Recall that $L$ is linear in $\lambda$, thus for the potential $u,\overline u$ which vanishes at $\pm\infty$ the boundary conditions for the eigenfunction $\Psi$ are written
\begin{align}
	\begin{split}
	\phi(x,\lambda) &\sim
	\begin{pmatrix}
		1\\
		0
	\end{pmatrix}e^{-i\lambda x},\quad 
	\overline \phi(x,\lambda) \sim
	\begin{pmatrix}
		0\\
		1
	\end{pmatrix}e^{i\lambda x}\quad \mathrm{as} \ x\to +\infty\\
	\psi(x,\lambda) &\sim
	\begin{pmatrix}
		0\\
		1
	\end{pmatrix}e^{i\lambda x},\quad 
	\overline \psi(x,\lambda) \sim
	\begin{pmatrix}
		1\\
		0
	\end{pmatrix}e^{-i\lambda x}\quad \mathrm{as} \ x\to -\infty.
	\end{split}
	\label{bc-IST}
\end{align}
The transitions coefficients $a(\lambda),b(\lambda)$, between $-\infty$ and $\infty$ are defined by the relations
\begin{align}
	\begin{split}
	\phi(x,\lambda)&= b(\lambda)\psi(x,\lambda) + a(\lambda) \overline \psi(x,\lambda)\\
	\overline \phi(x,\lambda) &= \overline a(\lambda) \psi(x,\lambda) + \overline b(\lambda) \overline \psi(x,\lambda).
	\end{split}
	\label{ab}
\end{align}
We skip the explanation of the inverse transform as it is exactly the standard theory where one can compute soliton solution out of the scattering data composed of $a,b$ and the associated eigenvalues $\lambda$. 
As we said before the time evolution of $a$ and $b$ are now stochastic and will computed in the following . 

We first introduce $M_\infty= M(x=\infty)$ and $\Xi_\infty = \Xi(x=\infty)$ which are matrices depending only on $\lambda$ and of the form $M_\infty= \mathrm{diag}(m_\infty,-m_\infty)$ and $\Xi_\infty= \mathrm{diag}(\xi_\infty,-\xi_\infty)$. 
A straightforward solution of \eqref{scatt-t} at $x=\pm\infty$ would be
\begin{align}
	v^+= 
	\begin{pmatrix}
		e^{m_\infty t + \xi_\infty W_t}\\
		0
	\end{pmatrix}\quad \mathrm{and}\quad 
	v^-= 
	\begin{pmatrix}
		0\\
		e^{-m_\infty t - \xi_\infty W_t}
	\end{pmatrix},
\end{align}
but would not be compatible with the fixed boundary conditions for $\phi$ and $\psi$ we defined in \eqref{bc-IST}. 
By defining intermediate functions one can arrive to an equation for the time evolution of the eigenfunctions $\phi$ and $\Psi$  
\begin{align}
	\begin{split}
	d\phi &= (M-M_\infty)\phi dt +( \Xi-\Xi_\infty)\phi \circ dW_t\\
	d\overline \phi &= (M+M_\infty)\overline \phi dt +( \Xi+\Xi_\infty)\overline \phi \circ dW_t.
	\end{split}
\end{align}
Written in term of $a$ and $b$ in \eqref{ab} we have $da= 0$ and $db = -2m_\infty bdt - 2\xi_\infty \circ dW$. 
It also turns out that the norming constants $c$ used in the construction of the soliton which decay rapidly at $\pm \infty$ is $c= b$. We will then denote
\begin{equation}
	b(\lambda, t) = c(t,\lambda )= e^{-2\Theta(t,\lambda)} b(\lambda,0)
\end{equation}
for an initial condition $b(\lambda,0)$ and where we defined
\begin{equation}
	\Theta:= m_\infty t+\xi_\infty W_t.	
	\label{Theta}
\end{equation}
Note that the diagonal coefficients $m_\infty$ and $\xi_\infty$ will be of the form $i\lambda^i$ where $i$ is the index of the flow. 
Then with the reduction $q=u, r=-\overline u$, the standard theory of IST gives the one soliton solution
\begin{equation}
	u(x,t) = -i\frac{\overline c(t)}{|c(t)|} 2\eta e^{-2i\xi x} \mathrm{sech}2(\eta x- \phi)
	\label{IST-sol}
\end{equation}
where $\phi$ is a phase such that $|c(0)|/(2\eta)= e^{2\phi}$ and the complex eigenvalue decomposes as $\lambda =\eta+ i\xi$ to give two real valued parameters $\eta$ and $\xi$.
After substituting $c(t)$ in Eq.~\eqref{IST-sol} and removing the extra constant phases we arrive at
\begin{align}
	u(x,t)= 2\eta e^{-2i\xi x}e^{2i\mathrm{Im}\,\Theta(t)} \mathrm{sech}[2\eta x + 2\mathrm{Re}\,\Theta(t) ].	
	\label{sech-sol}
\end{align}
Now, depending on the value of $\Theta$, namely depending on the choice of the stochastic time and flow time, we have the soliton solutions for the integrable stochastic PDEs. 
We will compute the explicit solution case by case in the next section. 

Following these ideas it is be possible to find other solutions, such as higher order solitons, or breathers. 
As we will see, the noise will affect differently the phase and the speed of the soliton. 
Hence, higher order solitons where their shape evolve periodically in time will have a more complicated stochastic behaviours, see Fig.~\ref{fig:collision}.

\subsection{Asymptotic behaviours} 
Following \cite{wadati1983stochastic} one can study the asymptotic behaviour of the expectation of a soliton for large times. 
Wadati found a diffusion of solitons where the width of the soliton increased $\propto t^{2/3}$ for large enough times. 
This analysis is based on a specific expansion of the hyperbolic secant function and one can easily see that his analysis still holds here for every case solitons found with the stochastic IST. 
The differences with the stochastic KdV equation are in the amplitude of the noise, which will depend not only on the deterministic speed but also on the shape of the soliton.
Reaching the asymptotic behaviour will take different times depending of the chosen parameters of the soliton. 
\section{Stochastic integrable equations}\label{SPDE}

We now turn into the study of the integrable stochastic equations of table \ref{tab:classification}. 
Instead of a detailed analysis for each equation, we will rather present a general survey of the principal features of these equations. 

\subsection{Stochastic KdV and NLS equations of type I}\label{SNLSI-section}
The first stochastic deformation of the KdV and NLS equations is with $k=1$, a stochastic time which corresponds to the travelling wave flow of the hierarchy. 
One easily obtain the following SPDEs
\begin{align}
	du &= i\left (u_{xx} + 2u|u|^2\right )dt + u_x\circ dW_t\label{SNLSI}\quad \mathrm{and}\\
	du &= \left (u_{xxx}+6u^{n+1}u_x\right )dt + u_x \circ dW_t\quad n=0,1,\label{SKdVI}
\end{align}
and their It\^o formulation 
\begin{align}
	du &= i\left (u_{xx} +2 u|u|^2\right )dt +\frac12 u_{xx}dt+ u_xdW_t\quad \mathrm{and}\\
	du &= \left (u_{xxx}+6u^{n+1}u_x\right )dt +\frac12 u_{xx}dt + u_x dW_t\quad n=0,1.
\end{align}
These equations can be transformed into their deterministic counterparts, via the following change of variable 
\begin{equation}
	x\to x-W_t. 
\end{equation}
The soliton solutions of both equations are then of the form $a(t,x-W_t)$ where $a(t,x)$ is the soliton of the deterministic equation. 
This transformation is natural because the stochastic time corresponds to the travelling wave equation $u_t = u_x$. 
Another explanation is that the stochastic time is the same as the space variable, thus a simple linear stochastic change of variable in the slice $N_{1i}$ for a given $i$ remove the noise of the PDE. 

\textit{One soliton solution from IST:}
With our choice of time and stochastic time we have for $\Theta$ defined in \eqref{Theta}  
\begin{equation}
	\Theta(t) = i\lambda^2 t + i\lambda W_t  = i(\xi^2-\eta^2+2i\eta\xi) t+i(\xi+i\eta)W_t.
\end{equation}
From Eq.~\eqref{sech-sol} the solution is given by
\begin{align}
	u(x,t)= 2\eta e^{-2i\xi x}e^{2i(\xi^2-\eta^2) t+2i\xi W_t )} \mathrm{sech}[2\eta (x - 2\xi t- W_t) ],
\end{align}
and corresponds to our solution with a stochastic spacial shift as derived before.

This stochastic KdV equation corresponds to the first instance of a stochastic integrable equation which was discovered and studied by \cite{wadati1983stochastic}. 
The correspondence of \eqref{SKdVI} with Wadati's stochastic KdV equation can be seen after a single change of variables.  
Indeed, Wadati started, for a Brownian motion $Z_t$,  with $u_t = (u_{xxx} + 6uu_x)dt  + d Z_t$ and, after the change of variable $u\to u- Z_t$ arrived at equation \eqref{SKdVI}, with $Z_t= dW_t$.
The latter identification is where Wadati's equation is different from equation \eqref{SKdVI} as he has one more degree of regularity in the noise compared to what we have here. 
We thus cannot directly make the link to its original equation, unless we assume a different type of noise in our theory.

Note that the stochastic NLS cannot be turned into an additive noise as for the KdV equation of Wadati as the modification of the amplitude of $u$ changes the form of the equation. 
This can also be seen from the fact that for the NLS equation the speed of a soliton is not linked to the soliton amplitude, thus the stochastic speed of the soliton cannot be converted to a stochastic amplitude. 

The physical interpretation of these equations is simple as the noisy term corresponds to having a small linear dispersion which fluctuate around its zero mean. 
Such system has already been studied by \cite{debouard2010nonlinear} where they instead considered a fluctuating second order dispersion. 
The fact that we can perform a change a variable to remove the first order dispersion makes this equation less interesting than the one of \cite{debouard2010nonlinear}, but if one uses a space-time noise $dW_t(x)$ then the change of variable is no more possible and the stochastic equation will have a fluctuating space dependent first order dispersion parameter. 
This stochastic PDE  could be physically relevant as the linear dispersion parameter is a measurable and important  quantity in the study of optical fibres for example, see \cite{agrawal2007nonlinear}.

\subsection{Stochastic NLS equation of type II and KdV/mKdV equations of type III}\label{SNLSII-section}

The second type of stochastic deformation of the NLS or KdV equations is when the flow time and the stochastic time are the same. 
They correspond to the SNLS of type II and SKdV of type III, as our classification is based on the stochastic time which is different in both cases. 
They read
\begin{align}
	du &= i\left (u_{xx} + 2u|u|^2\right )\left (dt +\circ dW_t\right )\label{SNLSII}\quad \mathrm{and}\\
	du &= \left (u_{xxx}+ 6u^{n+1}u_x\right )\left (dt  +	\circ dW_t\right ),\quad n=0,1.\label{SKdVII}
\end{align}
Written in this form it is clear that the effect of the noise can be understood as a stochastic parameter proportional to the right hand side of the equation, with mean $1$ and deviation given by the deviation of $dW_t$.  
The change of variable $t\to t+W_t$ will therefore map these equations into their deterministic counterparts. 
As before, this change of variable can be easily understood because the stochastic time is the same as the flow time. 
Note also that their It\^o correction is rather involved and not informative
\begin{align*}
	- u\left ( \overline u u_{xx}- u\overline u_{xx}+ |u|^4\right ) - |u|^2u_{xx} - \frac12 u_{xxxx} -  (u|u|^2)_{xx},
\end{align*}
thus we will not study it here.

\textit{One soliton solution from IST:}
We have for $\Theta$ in \eqref{Theta}
\begin{equation}
	\Theta (t) = i\lambda^2 t + i\lambda^2 W_t  = i(\xi^2-\eta^2+i2\eta\xi) (t+W_t),
\end{equation}
thus the speed of the soliton is given by $c(t) = 2\xi(1 + \dot W_t)$ which corresponds to the previous change of variables.  

\begin{figure}[htpb]
	\centering
	\includegraphics[width=0.6\textwidth]{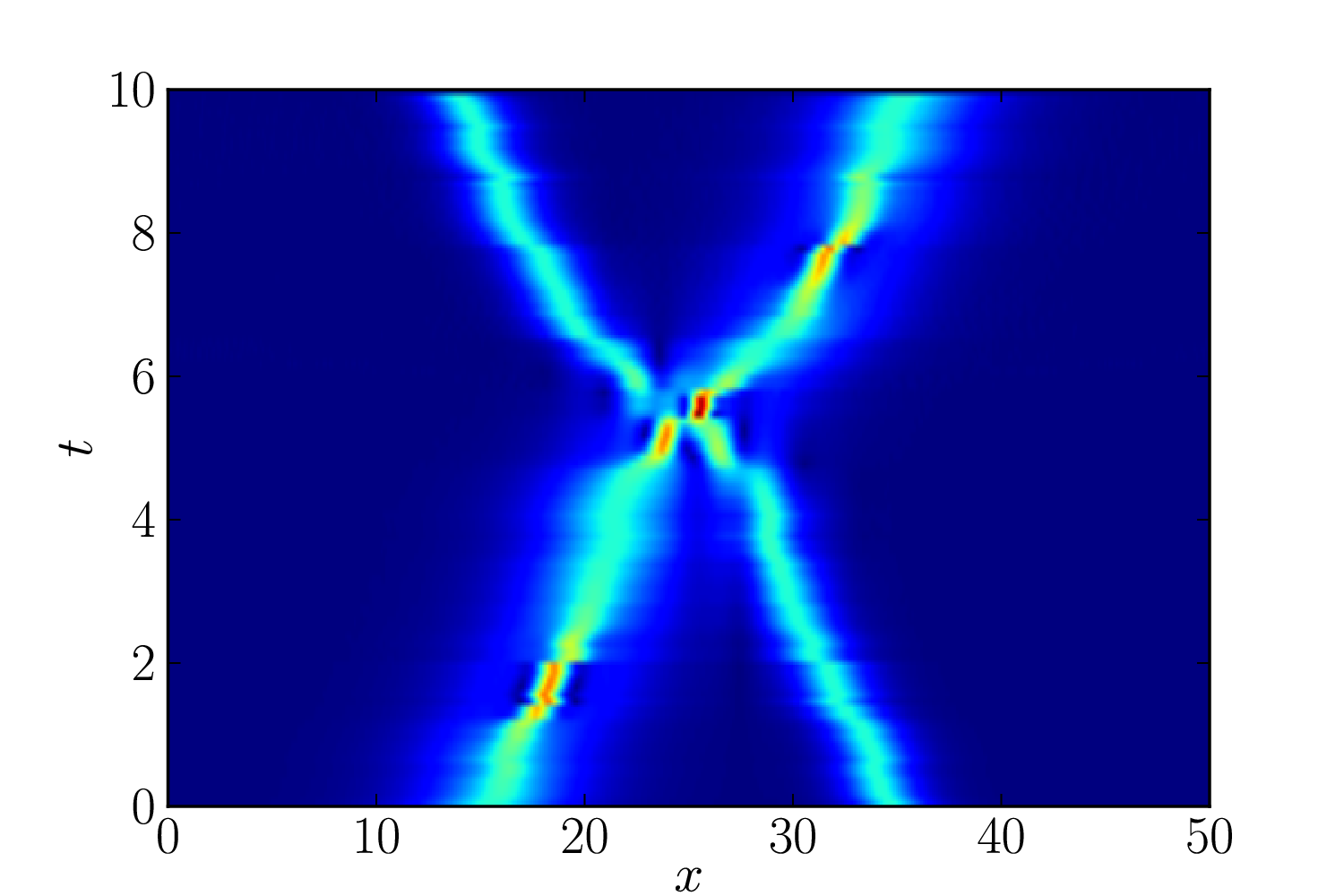}
	\caption{We display the amplitude $|u|$ of a numerical simulation of the stochastic NLS of type II with the following parameters: $\sigma=0.3$, $dt= 0.5\cdot 10^{-4}$ and $dx= 0.1$. 
		The initial condition consists of a soliton with $\eta = 0.5$ and $\xi=-0.5$ together with a second order soliton (the amplitude is multiplied by $2$ compared to the first order soliton) with $\eta= 0.25$ and $\xi= 0.5$. 
		This simulation illustrates several effects of the noise in the NLS equation.  
	First, the solitons keep their identity even after a collision. 
	Second, the phase shift of the solitons caused by the collision is not obvious to find as it will depend on the realisation of the noise.
Finally, the higher order soliton which is not a travelling wave sees its periodic motion affected by the noise in such a way that the motion remains cyclic, but with a random periodicity. }
	\label{fig:collision}
\end{figure}

This equation is thus not as simple as the type I equations as the noise amplitude is proportional to the speed of the deterministic soliton parametrised by $\xi$. 
From this fact, one may want to consider more complicated solutions, such as multi-solitons, higher order solitons, breathers, etc\dots
The noise would then influence the speed of the soliton, its phase shift as well as the speed of the oscillations of its shape when looking at higher order solitons or breathers for example. 
We illustrates these facts in Fig.~\ref{fig:collision} where we display the result of a numerical simulation of the collision of a first order and a second order soliton. 

\begin{figure}[htpb]
	\centering
	\includegraphics[width=0.9\textwidth]{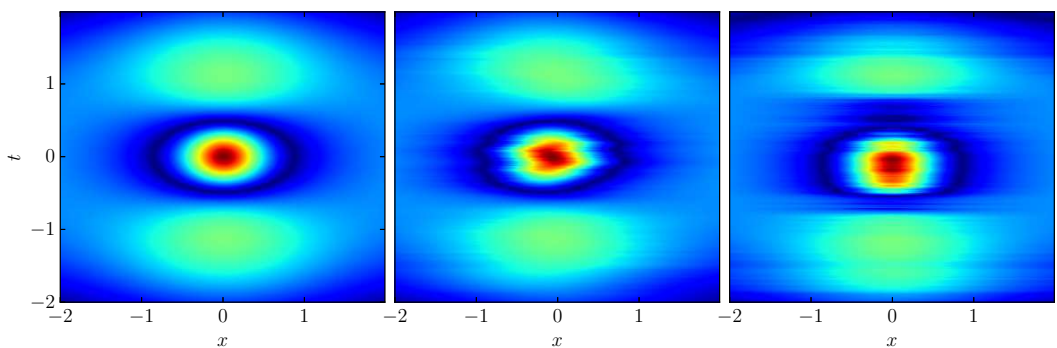}
	\caption{The Peregrine solutions of NLS and stochastic NLS equations are displays in the figure. 
	The left panel shows the deterministic breather, the middle panel shows the SNLS I and the right panel SNLS II. 
The effect of the noise is clearly different even with the same realisation and noise standard deviation of $\sigma=0.01$. }
	\label{fig:peregrine}
\end{figure}

\textit{Stochastic Peregrine solution:}
From the change of variables of the SNLS I and II, it is possible to find other solutions such that breather solutions of the NLS equation. 
We study here the classic Peregrine solution, given by
\begin{align}
	u(x,t)= \left ( 1- \frac {4(1+2it)}{1+ 4 x^2 + 4t^2}\right ) e^{it}.
\end{align}
Then, the stochastic solutions are given by $u(x+W_t,t)$ for the SNLS I and $u(x,t+W_t)$ for the SNLS II. 
We display in Fig.~\ref{fig:peregrine} the classical peregrine solution and the two stochastic peregrine solutions.
Although the realisation of the noise is the same for the two pictures, its influence on the peregrine solution is rather different. 
The SNLS I corresponds to a random horizontal shear of the solution whereas the SNLS II has a vertical stretching effect. 

\subsection{Stochastic complex mKdV of type II}\label{SKdVII-section}

Apart from the previous equations there is the intermediate choice $k=2$ for the stochastic time when the flow time is $t_3$.
In this case the reduction of the general AKNS hierarchy cannot be done with the KdV or even the mKdV variables but only with the NLS variables. 
Indeed, only the NLS reduction makes sense on the NLS flow with time $t_2$. 
The deterministic equation becomes a complex mKdV equation, known in the literature as the CmKdV of type II, or the Hirota equation first proposed in \cite{hirota1973exact}.
The noise is thus proportional to the NLS equation and the stochastic PDE reads
\begin{align}
	du = \left (u_{xxx}+ 6|u|^2u_x \right )dt +i\left (u_{xx} +2 u|u|^2\right )\circ dW_t.
\end{align}
There is no obvious change of variables which can remove the noise for this equation as in the previous cases. 
We can then just rely on the solutions given by stochastic IST. 

\textit{One soliton solution from IST:}
We have for $\Theta$ in \eqref{Theta}
\begin{align*}
	\Theta (t) &= i\lambda^3 t + i\lambda^2 W_t \\
	&=\left [ i( (\xi^2-\eta^2)\xi -2\eta^2\xi)-\eta( 3\xi^2-\eta^2)\right] t+\left [-2\eta\xi + i(\xi^2-\eta^2)\right]W_t. 
\end{align*}
Note that there is no obvious transformation that can remove the stochastic term in order to recast this equation into the deterministic CmKdV equation.  
The speed of the soliton is then given by $c(t)= ( 3\xi^2-\eta^2)+2\xi \dot W_t$, where the martingale part is proportional to $\xi$, the phase parameter of the solution. 
Therefore solitons with different values of $\xi$ will have a different noise amplitude. 
We refer to the discussion in the next section for the possible physical interpretation of this equation as they have the similar feature of mixing different flows.

\subsection{Stochastic NLS equation of type III}\label{SNLSIII-section}

The equation studied in this section is similar to the previous one, but with the time and stochastic time exchanged 
\begin{align}
	du = i\left (u_{xx} + 2u|u|^2\right )dt + \left (u_{xxx}+ 6u_x|u|^2 \right )\circ dW_t. 
\end{align}
In this case the noise is proportional to the CmKdV equation and the deterministic part is still the NLS equation. 

\textit{One soliton solution from IST:}
We have for $\Theta$ in \eqref{Theta}
\begin{align*}
	\Theta (t) &= i\lambda^2 t + i\lambda^3 W_t \\
	&=  \left [ i (\xi^2-\eta^2)-2\eta\xi\right] t +\left [ - \eta(3\xi^2 -\eta^2)+i((\xi^2-\eta^2)\xi-2\eta^2\xi)\right] W_t.
\end{align*}
As in the previous case, no obvious change of variable can transform this equation into its deterministic version. 
The speed of the soliton is then given by $c(t) =2\xi +[3\xi^2 -\eta^2] \dot W_t$.
Compared to the previous case, the noise and drift amplitude are reversed, thus the noise is proportional to $\xi^2$, and can even vanish for $\eta=\sqrt{3} \xi$ while having a stochastic phase. 

This equation might have physical interpretation, as the higher order noisy terms are sometimes included in the NLS equation for modelling higher order effects, for example in optical fibers (see \cite{agrawal2007nonlinear}). 
The noise that we have would model the fact that these higher order effects are randomly small. 
For example a recent study on a randomly small dispersion had been carried out in \cite{debouard2010nonlinear}. 

\textit{Stochastic Peregrine solution:}
As noted before this equation corresponds to the Hirota equation with a stochastic parameters in front of the mKdV terms. 
Remarkably, the Peregrine solution of the Hirota equation corresponds to the Peregrine solution of the NLS equation with a shift in the position of the form $x\to x+\gamma t$ if $\gamma$ is the coefficient in front of the mKdV terms. 
In our case of stochastic equations, replacing $\gamma t$ by $W_t$ gives a solution of the stochastic NLS of type III. 
Note that it is exactly the solution of the SNLS of type I but because the SNLSIII equation cannot be turned into the NLS equation this feature is just explained by a degeneracy of this solution.
Other solutions could be considered, such as higher order breathers of the Hirota equation where the noise would have a more complicated effect, see for example \cite{ankiewicz2010rogue}.  

\subsection{The stochastic CH equations}

In \cite{arnaudon2015lagrangian} the same procedure was used to deform the AKNS hierarchy and derive equations such that the Camassa-Holm equation \cite{camassa1993}. 
We can of course mix these two deformations and obtain stochastic equations of CH type.
We briefly look at a few of them here. 

Note that a change of Galilean frame can remove the third order dispersions of the CH or mCH equations to end up with dispersionless stochastic CH equations which admits peaked solutions known as peakons. 
We will show that when this is possible, the stochastic deformations of these equations reduces to a finite dimensional stochastic dynamical systems of peakons. 
Remark that the method we used to introduce the noise is similar to \cite{holm2015soliton}, where they introduced noise directly into the peakon system. 
As we will show, the stochastic CH equations derived here perfectly fit their framework.

We now make some comments on the equations coming from both the stochastic and Sobolev deformation of the AKNS hierarchy.

\textit{Stochastic CH equation of type I:}  The stochastic CH equation of type I corresponds to the same random shift in the space variable as for the stochastic KdV equation of type I.
This equation is a simplified version of the stochastic EPDiff equation of \cite{holm2015soliton} in the case where the arbitrary cylindrical noise they used is constant in space.
The reduction to a stochastic peakon system is possible but will not bring more informations about this equation.  

\textit{Stochastic CH equation of type II:}
The stochastic CH equation of type II is maybe the most interesting as it also reduces to a stochastic peakon dynamics but with a more interesting stochastic effects. 
Indeed, the stochastic CH equation of type II is given by
\begin{equation}
	dm = \left (2mu_x + m_x u \right )\left (dt +\circ dW\right ), \quad m=u-\alpha^2u_{xx}. 
\end{equation}
With the usual peakon Ansatz $u(x,t) = \sum_i p_ie^{-|x-q_i(t)|/\alpha}$ the variables $q_i(t)$ and $p_i(t)$ satisfy the following stochastic Hamiltonian system
\begin{align}
\begin{split}
	d q_i &= u(q_i,t)(dt +\circ dW)\\
	dp_i &= -p_iu(q_i,t) (dt +\circ dW). 
\end{split}
\end{align}
The Hamiltonian for this system is formally given by 
\begin{equation}
	h(p,q)dt = \frac12 \sum_{i,j} p_ip_je^{-|q_i-q_j|/\alpha} (dt +\circ dW),
\end{equation}
where the stochastic potential is the same as the deterministic Hamiltonian. 
This may be the only interesting stochastic peakon system which directly comes from the solution of a stochastic CH equation. 
Indeed, the higher flows are non local, thus we cannot easily use them as candidates for stochastic potentials. 
The deformation of the AKNS hierarchy of \cite{arnaudon2015lagrangian} does not take into account the negative flows of the dispersionless CH equation therefore from this viewpoint there is no other possibilities of adding noise to the CH equation while keeping its complete integrability.  

Another question for this equation, which could just be numerically studied in \cite{holm2015soliton} is whether the peakon crosses during an overtaking collision.  
It is straightforward to see that they will never cross by following the original proof of \cite{camassa1993}. 
Indeed, for the two peakons system the conserved quantities are $P= p_1+p_2$ and $H= \frac12 (p_1^2+ p_2)$, but if $q_1= q_2$, the Hamiltonian becomes $H= \frac12 (p_1+p_2)^2$ which contradicts its value when the peakons are separated. 
This proof works only if the Hamiltonian is a conserved quantity, which is the case here, but not in \cite{holm2015soliton}. 
This fact can also be seen directly from the two peakons solution, given in \cite{camassa1993} with the additional change of variable $t\to t+W_t$.

\textit{Stochastic CmCH equation of type III:} 
The Case III would correspond to the complex mCH equation of \cite{xia2015synthetical} with a noise proportional to the CH-NLS equation of \cite{arnaudon2015lagrangian,arnaudon2015deformation}. 
The integrability of the CH-NLS equation is not yet known as well as the physical interpretation of both equations. 
This stochastic equation is thus kept for future works.

\section{Conclusion and outlook}
We presented a systematic stochastic deformation of the AKNS hierarchy, based on the reformulation of this hierarchy in the framework of Lagrangian reduction by symmetries. 
We showed that the stochastic ZCR found by this method is equivalent to a stochastic IST. 
The latter was used to obtain one soliton solutions of the stochastic integrable equations in the case where the deterministic part is the NLS equation. 
The second part of this work was devoted to a survey of these new stochastic equations as well as the stochastic Camassa-Holm equation. 
The results of this section are just a beginning, and suggest different possible directions for future works. 
We list some of them here:
\begin{itemize}
	\item The stochastic IST could be applied to find other type of solutions, such as higher order solitons, other breathers or non solitary waves. 
		The noise will surely have a more complicated effect on the latter type of solutions. 
	\item We focused on the first stochastic times, but other times could be used. 
		In particular the time associated to the KdV5 equation together with the KdV equation for the deterministic part could be of interest. 
	\item Other hierarchies could also be considered, as this stochastic deformation method is rather general. 
	\item One can think of leaving the integrable world by letting the Brownian motion depend also on $x$, with some chosen spacial correlations. 
		If the spacial dependence stays small in some sense, we could expect that the integrability of the original equation would make this equation worth looking at.  
\end{itemize}

\subsection*{Acknowledgements}
		I am grateful to DD. Holm, AL. De Castro, AB. Cruzeiro, TS. Ratiu, X. Chen, S. Albeverio, N. Akhmediev and JM. Bismut for fruitful and thoughtful discussions during the course of this work. 
		I kindly acknowledge the Swiss National Science Foundation and the hospitality of the Centre Interfacultaire Bernoulli where this project emerged.  
	I acknowledge partial support from an Imperial College London Roth Award and from the European Research Council Advanced Grant 267382 FCCA.

\bibliographystyle{spmpsci.bst}
\bibliography{biblio.bib}

\end{document}